\documentclass{rspublic}
\usepackage{amssymb,amsmath}

\begin{document}

\title[Integrability and Linearization]{On the complete integrability and
linearization of nonlinear ordinary differential equations -
Part III: Coupled first order equations}

\author[Chandrasekar, Senthilvelan and Lakshmanan]{V. K. Chandrasekar,
M. Senthilvelan and M. Lakshmanan}

\affiliation{Centre for Nonlinear Dynamics, Department of Physics,
Bharathidasan Univeristy, Tiruchirapalli - 620 024, India}

\label{firstpage}

\maketitle

\begin{abstract}{Nonlinear differential equations, Coupled first order, Integrability, Integrating factor, Linearization}

\end{abstract}
Continuing our study on the complete integrability of nonlinear ordinary
differential equations, in this paper we consider the integrability
of a system of coupled first order nonlinear ordinary differential
equations (ODEs) of
both autonomous and non-autonomous types. For this purpose, we
modify the original Prelle-Singer procedure so as to apply it to
both autonomous and non-autonomous systems of coupled first order
ODEs. We briefly explain the method of finding integrals of motion
(time independent as well as time dependent integrals) for two and
three coupled first order ODEs by extending the Prelle-Singer(PS)
method. From this we try to answer some of the open questions in the
original PS method. We also identify integrable
cases for the two dimensional Lotka-Volterra system and
three-dimensional R$\ddot{o}$ssler system as well as other examples including
non-autonomous systems in a straightforward
way using this procedure. Finally, we develop a
linearization procedure for coupled first order ODEs.
\section{Introduction}
\label{sec4:1}

In our previous two works (Chandrasekar \textit{et al.} 2005; 2006)
we have studied in some detail the extended modified Prelle-Singer
(PS) procedure (Prelle \& Singer 1983; Duarte \textit{et al.} 2002)
so as to apply it
to a class of second and third order nonlinear ordinary differential
equations (ODEs) and solved several physically interesting nonlinear
systems and identified a number of important linearization
procedures. We now wish to extend the procedure to coupled ODEs. In
the present paper we discuss the modification and applicability of
the extended PS method to a system of first order ODEs of both
autonomous and non-autonomous types. In subsequent papers, we will
extend the procedure to coupled second and higher order ODEs. We are
motivated by certain open questions still prevailing in the original
PS method for a system of autonomous coupled first order ODEs. Before
discussing them and describing how we answer them, we shall have an
overview of the original PS method and its generalizations.

We recall that in 1983 Prelle \& Singer (1983) have proposed an
algorithmic procedure to find the integrating factor for the system
of two autonomous first order ODEs of the form
$\frac{dx}{dt}=P(x,y), \; \frac{dy}{dt}=Q(x,y)$, where $P$ and $Q$
are polynomials in $x$ and $y$ with coefficients in the field of
complex numbers. Equivalently it can be recast in the form $y' =
\frac{dy}{dx}=\frac{Q(x,y)}{P(x,y)}$. Once the integrating factor
for the latter equation is determined then it leads to a time
independent integral of motion for the above autonomous two coupled
first order ODEs (for the single first order ODEs the first integral
is nothing but the general solution). The PS method guarantees that
if the given system of two coupled first order ODEs has a first
integral in terms of elementary functions then this first integral
can be found. This method has been generalized to incorporate the
integrals with non-elementary functions (Singer 1990; Singer 1992;
Duarte \textit{et al.} 2002), and some first integrals of autonomous
systems of ODEs of higher dimensions (of dimension 3) were also
calculated. Man (1994) described a method for calculating first
integrals of autonomous systems which are rational or
quasi-rational, but said that 'The generalization of this procedure
to higher dimensions to find elementary first integrals is still an
open problem'. In addition, the question whether the PS procedure
can be extended to a non-autonomous system of first order ODEs has
not been addressed so far. Also to our knowledge the problem of
finding time dependent integrals for a given system of coupled first
order ODEs through this procedure has not been taken up so far.
Further, the problem of
finding both time dependent and independent integrals for a system
of first order ODEs greater than two in number has not been dealt
with systematically. In addition, the problem of how to linearize a
given system of coupled first order ODEs systematically also remains
to be tackled. In this paper, we address positively all these
questions and come out with valuable answers to these problems and
demonstrate the general results with suitable examples.

Firstly, in order to extend the PS method to a non-autonomous system
of first order ODEs with the rational form
$\frac{dx}{dt}=\frac{P_1(t,x,y)}{Q_1(t,x,y)}, \;
\frac{dy}{dt}=\frac{P_2(t,x,y)}{Q_2(t,x,y)}$, where $P_i(t,x,y)$ and
$Q_i(t,x,y),\;i=1,2,$ are analytic functions of $x$ and $y$ with
coefficients in the field of complex numbers, we develop a modified
technique and derive determining equations for the integrating
factors $R$ and $K$. In the case of a coupled system of two first
order ODEs we observe that the integrals of motion $I_1$ and $I_2$
of the coupled  ODEs are either both time dependent or $I_1$ may be
time independent while $I_2$ is time dependent (or vice versa).
Based on this
observation we analyze the problem by splitting it into two
different cases, namely $(i)$ the solutions of the determining
equations for $R$ and $K$ which lead to time independent integral(s)
and $(ii)$ the solutions which lead to time dependent integrals. We
show that the determining equations in the time independent integral
case coincide with the determining equations derived by Prelle and
Singer in their original paper (Prelle \& Singer 1983). In the
second case we obtain the time dependent integral, if it exists, and
from this we tackle the second problem. Using this method we find
integrable cases for the two dimensional Lotka-Volterra
(LV) equations (given as an example) with both time dependent and
independent integrals as well as certain non-autonomous systems.

For extending the method to a coupled system of more than two first
order ODEs, we first extend the above modified PS procedure to three
coupled first order ODEs and propose a systematic procedure to
obtain the integrating factors $R,\;K$ and $M$. In this case we
divide our analysis into five categories. We illustrate this theory
with two physically important examples, namely R$\ddot{o}$ssler
system and 3D LV system and identify some new integrable cases. The
method is extendable straightforwardly to a system of more than
three coupled first order ODEs. In addition to the above, we also
present a method of finding linearizing transformation for a
system of first order ODEs. We illustrate the theory with
certain concrete examples.

The paper is organized as follows.  In the following section we
develop the modified Prelle-Singer method applicable for
non-autonomous system of two coupled first order ODEs. In \S3, we
describe the method of solving the determining equations and  how
one can obtain compatible solutions from them. In \S4 we illustrate the
procedure with LV equation as an example and identify
many integrable cases in it. We also apply the method to
non-autonomous two coupled equations. In \S5, we  extend the PS
procedure to non-autonomous three coupled first order ODEs and
describe methods of solving the determining equations in \S6. We
emphasize the validity of the theory with two illustrative examples
arising  in different areas of physics in \S7. Further, in \S8 we 
discuss the direct applicability of the modified PS procedure to $n$ ($>$3) 
coupled first order ODEs. In \S9, we
demonstrate the method of identifying linearizing transformations
with examples. Finally, we present our conclusions in \S10.

\section{Prelle-Singer procedure for two coupled first order ODEs}
\label{sec4:2}
Let us consider a system of two coupled first order ODEs of the form
\begin{eqnarray}
\dot{x}=\frac{P_1(t,x,y)}{Q_1(t,x,y)}, \quad
\dot{y}=\frac{P_2(t,x,y)}{Q_2(t,x,y)}, \qquad (\;\;\dot{}=\frac{d}{dt})
\label {eq01}
\end{eqnarray}
where $P_i$ and $Q_i,\;i=1,2,$ are analytic functions of $x$ and $y$ with
coefficients in the field of complex numbers. Further, we assume that the
ODE (\ref{eq01})
admits a first integral $I(t,x,y)=C,$ with $C$ constant on the
solutions, so that the total differential becomes
\begin{eqnarray}
dI={I_t}{dt}+{I_{x}}{dx}+{I_{y}}{dy}=0, \label {eq02}
\end{eqnarray}
where subscript denotes partial differentiation with respect to that variable.
Let us rewrite ({\ref{eq01}) in the form
\begin{eqnarray}
\frac{P_1}{Q_1}dt-dx=0,\qquad
\frac{P_2}{Q_2}dt-dy=0. \label {eq03}
\end{eqnarray}
Hence, on the solutions, the 1-forms (\ref{eq02}) and
(\ref{eq03}) must be proportional. Multiplying the first equation in (\ref{eq03})
by the integrating factor $ R(t,x,y)$ and the second equation in (\ref{eq03}) by a second integrating
factor $ K(t,x,y)$ (both of which are to be determined), we have on the
solutions that
\begin{eqnarray}
&&dI=(R\phi_1+K\phi_2)dt-Rdx-Kdy=0,\;\;\label {eq04}
\end{eqnarray}
where $ \phi_i\equiv {P_i}/{Q_i},\; i=1,2$. Comparing equations (\ref{eq04})
and (\ref{eq02}) we have, on the solutions, the relations
\begin{eqnarray}
 I_t  =(R\phi_1+K\phi_2),
 \quad  I_{x}  = -R, \quad I_{y} =-K.
  \label {eq05}
\end{eqnarray}

The compatibility conditions, $I_{tx}=I_{xt}$, $I_{ty}=I_{yt}$,
$I_{xy}=I_{yx}$, between the equations
(\ref{eq05}) provide us the conditions
\begin{eqnarray}
R_t+\phi_1R_x+\phi_2R_y  &=&-(R\phi_{1x}+K\phi_{2x}),  \label {eq06}\\
K_t+\phi_1K_x+\phi_2K_y  &=&-(R\phi_{1y}+K\phi_{2y}), \label {eq07}\\
R_{y} &=&K_{x}. \label {eq08}
\end{eqnarray}

Integrating equations~(\ref{eq05}), we obtain the integral of motion,
\begin{eqnarray}
I=r_1+r_2
-\int\bigg[K+\frac{d}{dy}\bigg(r_1+r_2\bigg)\bigg]
dy,
\label {eq09}
\end{eqnarray}
where
\begin{eqnarray}
r_1&=\int(R\phi_1+K\phi_2)dt,
\qquad
r_2=-\int(R+\frac{d}{dx}(r_1))dx.
\nonumber
\end{eqnarray}
Solving the determining equations (\ref{eq06})-(\ref{eq08})
consistently, we can obtain expressions for the functions $R$ and
$K$. Substituting them into (\ref{eq09}) and evaluating the
integrals we can deduce the associated integral of motion. If two
independent sets of solutions $R$ and $K$ for
(\ref{eq06})-(\ref{eq08}) are found then they give rise two
independent integrals for the given system of first order ODEs
(\ref{eq01}) which ensures the complete integrability of the system and obtaining explicit general solution of the system.
\section{Method of Solving Determining Equations}
\label{sec4:3} One may note that the determining
equations~(\ref{eq06})-(\ref{eq08}) are over-determined and the crux
of the problem lies in finding the explicit solutions satisfying all
the three determining equations, since once a particular solution is
known then the integral of  motion can be readily constructed. To
solve the equations  (\ref{eq06}) and (\ref{eq07}) we introduce a
transformation
\begin{eqnarray}
R=SK, \label {eq09a}
\end{eqnarray}
where $S$ is a function of $t,x$ and $y$ so that the
determining equations (\ref{eq06})-(\ref{eq08}) now become
\begin{eqnarray}
S_t+\phi_1S_x+\phi_2S_y  &=&-\phi_{2x}+(\phi_{2y}-\phi_{1x})S+\phi_{1y}S^2,
\label {eq10}\\
K_t+\phi_1K_x+\phi_2K_y  &=&-K(S\phi_{1y}+\phi_{2y}), \label {eq11}\\
K_{x} &=&SK_{y}+KS_{y}. \label {eq12}
\end{eqnarray}
One may observe that the equation for $S$, namely equation
(\ref{eq10}), is decoupled from that of $K$ (equations (\ref{eq11})
and (\ref{eq12})) and so the set (\ref{eq10})-(\ref{eq12}) may be
easier to analyze than solving the original ones
(\ref{eq06})-(\ref{eq08}) directly.

To begin with we observe that for the system of two coupled first
order ODEs (\ref{eq01}) there can be two independent integrals $I_1$
and $I_2$ such that either both of them are time dependent or $I_1$
may be time independent while $I_2$ is time dependent (or vice versa). So we
consider these two cases separately with corresponding solutions $S$
and $K$ for equations (\ref{eq10})-(\ref{eq12}). The determining
equations in the time independent integral case (see equation
(\ref{eq15}) below) coincides with the determining equation derived
by Prelle and Singer in their original paper (Prelle \& Singer
1983), whereas for the time dependent integrals case we develop an
extended procedure to capture both the integrals, if they exist, and
thereby making the PS procedure a more powerful tool in a
self-contained way.

\subsection{Time independent integral}
\label{sec4:3:1}
In the case $I_t=0$ we denote $I_1=I$ and note that the function $S$ can be easily
fixed with
the help of the first equation in (\ref{eq05}), that is,
\begin{equation}
\frac{R}{K}=S= -\frac{\phi_{2}}{\phi_{1}}.
\label{eq12a}
\end{equation}
Since $I$ is independent of $t$, it follows from equation (\ref{eq05}) that $S$
(and also $R$ and $K$) is also independent of $t$.
Indeed one can check that $S=-\frac{\phi_{2}}{\phi_{1}}$ is a solution of
(\ref{eq10}). Now substituting $S=-\frac{\phi_{2}}{\phi_{1}}$ into
(\ref{eq11}) we get the following equation for $K$, that is,
\begin{eqnarray}
\phi_1K_x+\phi_2K_y  &=&K(\frac{\phi_{2}}{\phi_{1}}\phi_{1y}-\phi_{2y}).
\label {eq13}
\end{eqnarray}

We now make a substitution
\begin{equation}
K = \frac{\phi_{1}}{f(x,y)},
\label{eq14}
\end{equation}
where $f(x,y)$ is an arbitrary non-zero function of $x$ and $y$.
Then (\ref{eq13}) takes the simpler form
\begin{eqnarray}
\phi_1f_x+\phi_2f_y   &=&f(\phi_{1x}+\phi_{2y}). \label {eq15}
\end{eqnarray}
We note that by redefining $f=\frac{1}{\mathcal{R}}$
equation (\ref{eq15}) coincides with the
determining equation derived by Prelle \& Singer (1983) for the autonomous case
of (\ref{eq01}). We also mention that equation (\ref{eq15}) is nothing but
the one obtained by substituting the forms of $S$ and $K$ given in (\ref{eq12a})
and (\ref{eq14}) into the constraint equation (\ref{eq12}). So by solving equation
(\ref{eq15}) we can get the complete set $R(=SK)$ and $K$ associated with
equations  (\ref{eq10})-(\ref{eq12}).

Even though equation (\ref{eq15}) is a quasilinear PDE in two
variables, the associated characteristic equation again leads to
coupled differential equations of the form (\ref{eq01}). Thus the
routine methods of finding general solution of quasilinear PDEs
are not very useful
here. Here, we find particular solutions for
the determining equation (\ref{eq15}) in a different way and obtain
integrable cases for the given system. For this purpose, we
assume specific functional form for $f(x,y)$ with unknown functions
and determine the later consistently. A simple but effective choice
is $f=(A(x)+B(x)y)^r$, where $A$ and $B$ are functions of their
arguments, and $r$ is a constant. Again the reason for choosing this
form is as follows. Since $K$ is in a rational form, while taking
differentiation or integration the form of the denominator remains
the same but the power of the denominator decreases or increases by
a unit order from that of the initial one. So instead of considering
$f$ to be of the form $f=A(x)+B(x)y$, one may consider a more
general form $f=(A(x)+B(x)y)^r$, where $r$ is a constant to be
determined. Depending on the problem in hand, one can also assume more
general form and proceed as in the present case.

Substituting now the form $f=(A(x)+B(x)y)^r$ into (\ref{eq15}) we
arrive at the following equation for the unknown functions $A$ and
$B$, that is,
\begin{eqnarray}
r\bigg[\phi_{1}(A_x+B_xy)+\phi_{2}B\bigg]=(\phi_{1x}+\phi_{2y})(A+By).
\label{eq16}
\end{eqnarray}
Inserting the given form of $\phi_i$'s, $i=1,2,$ into the above
equation (\ref{eq16}) and solving the resultant equation, one can
fix the forms of $A,\;B$ and $r$. Now plugging the resultant form of
$f$ into equation~(\ref{eq14}) one can get the integrating factor
$K$ which in turn leads us to the other integrating factor $R$
through the relation (\ref{eq12a}). Finally, substituting $R$ and
$K$ into equation~(\ref{eq09}) and evaluating the integrals one can
deduce the time independent integral for the given system.
Since we are dealing a system of two first order ODEs, this time
independent integral itself guarantees the integrability of the given system.
However, to explore  the general solution one may seek the time dependent
second integral. We describe the procedure in the following sub-section.
\subsection{Time dependent integral}
\label{sec4:3:2} Now we focus our attention on the case $I_t\neq0$.
In this case, the function $S$ has to be determined from
equation~(\ref{eq10}). Since it is too difficult to solve
equation~(\ref{eq10}) for its general solution, we seek particular
solutions for $S$, which is sufficient for our purpose. In
particular, we seek a simple rational expression for $S$ in the form
\begin{eqnarray}
S = \frac{A_1(t,x)+B_1(t,x)y}{A_2(t,x)+B_2(t,x)y},
\label{eq18}
\end{eqnarray}
where $A_i$'s and $B_i$'s, $i=1,2,$ are arbitrary functions of $t$
and $x$ which are to be determined. Of course this can be further
generalized, if need arises. Substituting (\ref{eq18}) into
(\ref{eq10}) and equating the coefficients of different powers of
$y$ to zero, we get a set of determining equations for the functions
$A_i$'s and $B_i$'s, $i=1,2$. Solving these determining equations
we obtain explicit expressions for the functions $A_i$'s and $B_i$'s,
$i=1,2,$ which in turn fixes $S$ through the relation (\ref{eq18}).

Now substituting the forms of $S$ into equation~(\ref{eq11}) and solving the
resultant equation one can obtain the corresponding forms of $K$. To solve the
determining equation for $K$ we again seek the same form of ansatz (\ref{eq14})
but with explicit $t$ dependence on the coefficient functions, that is, $K =
\frac{S_d}{(A(t,x)+B(t,x)y)^r},$ where $S_d$ is the denominator of $S$. Once $S$
and $K$ are determined then one has to verify the compatibility of this set
$(S,K)$ with the extra constraint equation~(\ref{eq12}). Now substituting $R$'s
$(=SK)$ and $K$'s into equation~(\ref{eq09}) and evaluating the integrals one can
construct the associated integrals of motion.

We note here that for a given equation (\ref{eq01}) one may also get
two time dependent integrals or one time dependent and one time
independent independent integrals (discussed earlier) which in turn
automatically guarantees
the complete integrability of the given system and provide us explicit
solution by algebraic manipulation. On the other hand, under
certain circumstances, one may get only one time dependent integral
and one can transform this time dependent integral
into time independent one and thereby establish the integrability.

\section{Two coupled ODEs - Application}
\label{sec4:4}
\subsection{Example: Two dimensional Lotka-Volterra (LV) system}
\label{sec4:4:1}
Our motivation is to identify integrable cases and
deduce both time dependent and independent integrals for a given
system of two coupled first order ODEs  through the extended
Prelle-Singer procedure in a self-contained way.

To demonstrate this we consider the celebrated two dimensional LV
system
\begin{eqnarray}
\dot{x}=x(a_{1}+b_{11}x+b_{12}y)=\phi_{1},\quad
\dot{y}=y(a_{2}+b_{21}x+b_{22}y)=\phi_{2},
\label{2ex01}
\end{eqnarray}
where $a_i$'s and $b_{ij}$'s $i,j=1,2,$ are six real parameters.
This system was originally introduced by Lotka (Lotka 1920) and
Volterra (Volterra 1931) to model two species competition. However,
in recent years this model appears widely in applied mathematics and
in a large variety of physics topics such as laser physics, plasma
physics, convective instabilities, neural networks, etc. (Minorsky
1962; Brenig 1988; Murray 1989). The integrability properties of the
system (\ref{2ex01}) alone have been analyzed by many authors, see
for example Cairo \& Llibre (2000) and references therein.

In the following, we identify integrable cases in
(\ref{2ex01}), through our procedure.
\subsubsection{Time independent integral ($I_t=0$)}
\label{sec4:4:1:1}
In this case the function $S$ can be fixed easily in the form (vide equation
(\ref{eq12a}))
\begin{equation}
S=-\frac{\phi_{2}}{\phi_{1}}=-\frac{y(a_{2}+b_{21}x+b_{22}y)}{x(a_{1}
+b_{11}x+b_{12}y)}.
\label{2ex01a}
\end{equation}
To explore the integrating factor $K$ we need to fix the form $f$
first, see equation (\ref{eq14}). For this purpose we substitute the
$\phi_i$'s, $i=1,2,$ into (\ref{eq16}) so that one gets the
following equation for the unknown functions $A$ and $B$ which
constitute the function $f$, that is,
\begin{eqnarray}
&&r\bigg[x(a_{1}+b_{11}x+b_{12}y)(A_x+B_xy)
+y(a_{2}+b_{21}x+b_{22}y)B\bigg]\nonumber\\
 &&\qquad\qquad=(a_1+a_2+(b_{21}+2b_{11})x+(b_{12}+2b_{22})y)(A+By).
\label{2ex02}
\end{eqnarray}
Equating the coefficients of various powers of $y^{i}$, $i=0,1,2,$
and solving the resultant differential equations for $A$ and $B$ we
arrive at the following two general expressions which involve the
system parameters, that is,
\begin{eqnarray}
&& b_{21}(b_{12}-b_{22})(b_{12}b_{21}-b_{11}(b_{12}-2b_{22}))
(a_1(b_{11}-b_{21})b_{22}+a_2b_{11}(b_{22}-b_{12}))=0,\nonumber\\
&& a_1b_{22}(b_{11}-b_{21})^2(a_2b_{12}-a_1b_{22})
+a_2^2b_{11}(b_{12}-b_{22})(b_{12}b_{21}-b_{11}b_{22})=0, \label
{2ex04}
\end{eqnarray}
and
\begin{eqnarray}
r=\frac{(b_{11}b_{12}+b_{12}b_{21}-2b_{11}b_{22})}{b_{11}(b_{12}-b_{22})},
\quad \mbox{or} \quad
r=\frac{(b_{11}b_{12}+b_{12}b_{21}-2b_{11}b_{22})}{(b_{12}b_{21}-b_{11}b_{22})}.
\label {2ex04aa}
\end{eqnarray}
Any consistent solution which comes out from the above expression (\ref{2ex04})
gives us an integrating factor
which in turn leads us to an integral of motion. In this sense (\ref{2ex04}) forms
an integrability condition of some generality which in fact encompasses all known integrable cases with time independent integrals (Cairo \& Llibre 2000; Llibre \& Valls 2007). For example, while
analysing the above equation, we find that one can straightforwardly recover several known integrable cases like $(i)\;b_{11}=b_{22}=0,r=1,
\;(ii)\;b_{22}=b_{12},b_{11}=b_{21},r=1,\;(iii)\;a_1 = a_2, b_{12} = 3b_{22}, b_{11} = -b_{21},r=-1,\;(iv)\;a_1=a_2=0,\;r$ as given in equation (\ref{2ex04aa}), and so on
straightforwardly from equations (\ref{2ex04}) and (\ref{2ex04aa}) and construct the associated  integral of motion which in turn coincide  with the existing results.  However, as we are interested to construct an
integral of motion with more general parametric choice we do not fix
any relation between parameters (other then the general relation) and
proceed further.


The respective forms of $A$ and $B$ are (for $ b_{22}b_{12}(b_{22}-b_{12})\neq0$),
\begin{eqnarray}
A=((r-1)g(x)+rb_{22}(a_1+b_{11}x))x^{\frac{((2-r)b_{22}+b_{12})}{rb_{12}}},\;
B=rb_{22}b_{12}x^{\frac{((2-r)b_{22}+b_{12})}{rb_{12}}}
\label{2ex02a}
\end{eqnarray}
so that
\begin{eqnarray}
f&=&x^{\frac{((2-r)b_{22}+b_{12})}{b_{12}}}\bigg((r-1)g(x)+rb_{22}(a_1+b_{11}x
+b_{12}y)\bigg)^{r}, \label{2ex03}
\end{eqnarray}
where
\begin{eqnarray}
g(x)=\bigg(a_2b_{12}-2a_1b_{22}
+\frac{b_{22}}{(b_{22}-b_{12})}(b_{12}b_{21}+b_{11}(b_{12}-2b_{22}))x\bigg).\label{gandx}
\end{eqnarray}

Making use of the explicit forms of $S$ and $f$, vide equations~ (\ref{2ex01a}) and
(\ref{2ex03}) respectively, with the parametric restrictions (\ref{2ex04}),
the integrating factors, $K$ and $R$, can
be fixed as $R=-\frac{\phi_2}{f}$ and $K=\frac{\phi_1}{f}$.
Substituting the forms $R$ and $K$ into (\ref{eq09}) and evaluating
the  integrals we arrive at the following time independent integrals
for (\ref{2ex01}) for the parametric cases (\ref{2ex04}) for
different values of $r$, namely
\begin{eqnarray}
&&I=\frac{x}{f}\bigg[\frac{(r-1)}{r^2}g(x)^2
   +\bigg[b_{22}(a_1+b_{11}x+b_{12}y)(a_2b_{12}-a_1b_{22}
\label{2ex06a}\\
  &&\qquad +\frac{b_{22}}{(b_{12}-b_{22})}(b_{11}b_{22}-b_{12}b_{21})x+b_{12}b_{22}y))\bigg]\bigg],\qquad
\quad \; r\neq0,2,
\nonumber\\
&&I=
\frac{2b_{22}(a_1+b_{11}x+b_{12}y)}
{b_{12}h(x,y)}-\log\bigg[x^{-\frac{b_{22}}{b_{12}}}h(x,y)\bigg],\;\;\;\;
r=2,\label{2ex06b}\\
&&I=x^{-\frac{2b_{22}}{b_{12}}}y(2a_1+2b_{11}x+b_{12}y),\qquad
\qquad \qquad\qquad
 r=0,
\label {2ex06c}
\end{eqnarray}
where $h(x,y)=a_2+(b_{11}+b_{21})x+2b_{22}y$ and $g(x)$ is given in
(\ref{gandx}).

For $ b_{22}b_{12}(b_{22}-b_{12})=0$, one obtains known integrable
cases following the same procedure as above. For example, let us
consider the first case, which we cited above as the known case
$b_{22}=b_{11}=0$. For this case, one can find a
trivial solution for equation (\ref{2ex02}) as $A=0$ and $B=x$
with $r=1$ so that $f$ becomes $f=xy$. The respective integrating
factors $R$ and $K$ read as $R=-\frac{a_2+b_{21}x}{x}$ and
$K=\frac{a_1+b_{12}y}{y}$  so that the associated integral of motion
takes the form $I=b_{21}x-b_{12}y+a_2\log{x}-a_1\log{y}$. This
integral is well known and popular in the literature for a long time
(see for example Minorsky 1962; Prelle \& Singer 1983; Murray 1989).
Similarly for the case $b_{22}=b_{12}$ and $b_{11}=b_{21}$, we obtain
the integral of motion of the form $I=x^{a_2}y^{-a_1}
(a_1a_2+a_1b_{22}y+a_2b_{11}x)^{(a_1-a_2)}$.
In the case $b_{12}=0$, leads to the uncoupled equation for LV and
the general solution for this case can easily be obtained.

\subsubsection{Time dependent integrals ($I_t\neq0$)}
\label{sec4:4:1:2}
Now let us concentrate on  the case $I_t\neq0$. In this case $S$
has to be determined from equation~(\ref{eq10}), that is,
\begin{eqnarray}
S_t+x(a_{1}+b_{11}x+b_{12}y)S_x+y(a_{2}+b_{21}x+b_{22}y) S_{y}
=-b_{21}y
\nonumber\\\qquad \qquad\qquad+((a_{2}+b_{21}x+2b_{22}y)-(a_{1}+2b_{11}x+b_{12}y))S
+b_{12}xS^2.
   \label{3ex01}
\end{eqnarray}
As we mentioned earlier, to obtain a particular solution for the
above equation (\ref{3ex01}) we seek a simple ansatz for $S$ of the
form (\ref{eq18}). Substituting (\ref{eq18}) into (\ref{3ex01}) and
solving the resulting equation we obtain non-trivial forms of $S$
for the following specific parametric  restrictions (we omitted the
uncoupled case $b_{12}\;b_{21}=0$), namely
\begin{eqnarray}
  (i)\;&&a_1=a_2,\;\;b_{21}-b_{11}(2-\frac{b_{12}}{b_{22}})=0,\label{3ex02a}\\
  (ii)\;&&b_{21}=b_{11},\;\;b_{12}=b_{22},\label{3ex02b}\\
  (iii)\;&&b_{21}=\frac{b_{22}b_{11}}{b_{12}},\label{3ex02c}\\
  (iv)\;&&a_1=a_2,\;\;b_{21}-b_{11}(2-\frac{b_{12}}{b_{22}})\neq0,
\label{3ex02d}
\end{eqnarray}
and the respective $S$ forms are
\begin{eqnarray}
   (i)&& S=\frac{b_{11}}{b_{22}},\qquad
  (ii)\;S=-\frac{y}{x},\nonumber\\
  (iii)&&S=-\frac{b_{22}y}{b_{12}x},\quad
  (iv)\;S=-\frac{y(b_{21}x+b_{22}y)}{x(b_{11}x+b_{12}y)}.
\label{3ex03}
\end{eqnarray}
Now substituting the above forms of $S$ into equation~(\ref{eq11})
and solving the resultant equation we obtain the corresponding forms
of $K$. By making use of the ansatz mentioned in \S3$\;b$ we obtain following expressions for $K$, namely
\begin{eqnarray}
  (i)\;\;&& K=-\frac{a_2b_{22}e^{-a_2 t}}{(a_2+b_{11}x+b_{22}y)^2},\qquad
  (ii)\;\;K=\frac{x}{y^2}e^{(a_2-a_1)t},\nonumber\\
  (iii)\;\;&&K=-x^{-\frac{b_{22}}{b_{12}}}e^{(\frac{a_1b_{22}}{b_{12}}-a_2)t},\nonumber\\
  (iv)\;\;&&K=\frac{(b_{11}x+b_{12}y)e^{\frac{a_2(b_{12}-b_{22})}{b_{12}}
(r-2)t}x^{\frac{(r-2)b_{22}}{b_{12}}}}
{\bigg(\frac{(r-1)}{r}\frac{(b_{12}b_{21}+b_{11}(b_{12}-2b_{22}))}
{(b_{12}-b_{22})x}-b_{11}x-b_{12}y\bigg)^{r}}, \label{3ex04}
\end{eqnarray}
with $r$ is given in equation ({\ref{2ex04aa}). It may be noted that the set
(\ref{3ex02a})-(\ref{3ex02d}) also includes the known time dependent integrable cases.

Once $R(=SK)$ and $K$ are determined then one has to verify the compatibility
of this solution with the extra constraint (\ref{eq08}) which indeed gets
satisfied in each one of the above four cases.
Substituting the resultant integrating factors into (\ref{eq09}) and
evaluating the integrals we obtain the associated time dependent
integrals of motion in the forms
\begin{eqnarray}
   (ia)\;&& I=\frac{e^{-a_2 t}(b_{11}x+b_{22}y)}{(a_2+b_{11}x+b_{22}y)},\;\;
  a_2\neq0,
\label{3ex05a}\\
  (ib)\;&&I=\frac{1+(b_{11}x+b_{22}y)t}{(b_{11}x+b_{22}y)},\;\;
  a_2=0,
\label{3ex05b}\\
  (ii)\;&&I=\frac{x}{y}e^{(a_2-a_1)t},\label{3ex06}\\
  (iii)\;&& I=e^{(\frac{a_1b_{22}}{b_{12}}-a_2)t}x^{-\frac{b_{22}}{b_{12}}}y,
  \label{3ex07}
  \end{eqnarray}
\begin{eqnarray}
  (iva)\;&& I=
\bigg(\frac{(r-1)}{r}\frac{d_3x}{d_1}-e_2\bigg)^{-r}
\bigg[\frac{(r-1)}{r^2}d_3^2x^2+e_2d_1
\nonumber\\
  &&\quad\times\;
(b_{11}b_{22}x+b_{12}^2y-b_{12}e_1)\bigg]
  x^{\frac{(r-2)b_{22}}{b_{12}}}e^{\frac{a_2d_1}{b_{12}}(r-2)t},\;\;\;\;\;\; r\neq0,2,
\label{3ex08a}\\
  (ivb)\;&&  I=\log\bigg[2x^{-\frac{b_{22}}{b_{12}}}yd_1\bigg]
-\frac{b_{21}x}{b_{12}y}-\frac{a_2d_1}{b_{12}}t,\qquad
\qquad \qquad\quad\;\; r=2,
\label{3ex08b}\\
  (ivc)\;&& I=e^{-\frac{2a_2d_1}{b_{12}}t}
x^{-\frac{2b_{22}}{b_{12}}}\bigg(b_{11}^2d_2x^2
+b_{12}^2d_1y^2+b_{11}b_{12}e_4x\bigg),\; r=0,
\label{3ex08c}
\end{eqnarray}
where
\begin{eqnarray}
d_1&=&(b_{12}-b_{22}),\qquad \qquad \;\;\;\;e_1=(b_{21} x+b_{22}y),\;\;\;
d_2=(b_{12}-2b_{22}),\nonumber\\
d_3&=&(b_{12}b_{21}+b_{11}d_2),\qquad\;\;\;
e_2=(b_{11}x+b_{12}y),\nonumber\\
e_3&=&((b_{11}-b_{21})x+2d_1y),\;\;e_4=(b_{21}x+2d_1y).
\end{eqnarray}
Finally we note that our method not only gives us a rather general set of
integrable parametric relations (which includes all known cases) but
also provides two independent integrals from which one can deduce
the general solution for some cases.
\subsubsection{General solutions/Integrability}
\label{sec4:4:1:3}
Interestingly, we find that for certain
parametric choices we have two independent integrals (time
independent as well as time dependent) and consequently one can
express  the general solution explicitly by using both of them. For example,
let us consider the parametric choice given in (\ref{3ex02a}) and
the associated time dependent integrals given in (\ref{3ex05a}) and
(\ref{3ex05b}). For this parametric choice, that is,
$a_1=a_2,\;\;b_{21}b_{22}-b_{11}(2b_{22}-b_{12})=0$, we can also find
the following time independent integral from (\ref{2ex06a}),
\begin{eqnarray}
I_2&=&\frac{y}{x}(a_1+b_{11}x+b_{22}y)^{\frac{b_{12}}{b_{22}}-1}.
\label {3ex09}
\end{eqnarray}
Using the integrals $I$ and $I_2$, the general solution for the 2D
LV system, (\ref{2ex01}), for the parametric choice (\ref{3ex02a})
can be written as
\begin{eqnarray}
x(t)=\frac{a_1\hat{a}_1e^{a_1t}I}{b_{11}\hat{a}_1g_1
+b_{22}I_2(g_1)^{\frac{b_{12}}{b_{22}}}},\;
y(t)= \frac{a_1e^{a_1t}II_2}
{(b_{11}\hat{a}_1(g_1)^{2-\frac{b_{12}}{b_{22}}}
+b_{22}I_2g_1)},\; a_1\neq0
 \label {3ex11a}
\end{eqnarray}
and
\begin{eqnarray}
x(t)=\frac{1}{I_2(g_2)^{\frac{b_{12}}{b_{22}}}-b_{11}g_2},\;\;
y(t)=\frac{I_2(g_2)^{(\frac{b_{12}}{b_{22}}-1)}}{b_{22}(b_{11}g_2
 -I_2(g_2)^{\frac{b_{12}}{b_{22}}})},\;\; a_1=0,
\label{3ex11b}
\end{eqnarray}
where $\hat{a}_1=a_1^{\frac{b_{12}}{b_{22}}-1},\;g_1=(1-I e^{a_1t})$ and
$g_2=(t+I)$, respectively. Further, for the parametric choice given in
(\ref{3ex02b}), that is, $b_{21}=b_{11},
\;\;b_{12}=b_{22}$, we obtain the general solution of the form
\begin{eqnarray}
x(t)=\frac{a_1a_2e^{a_1t}}{(II_1^{-a_2})^{\frac{1}{a_1-a_2}}-g_3(t)},\;\;
y(t)=\frac{a_1a_2e^{a_2t}}{(II_1^{-a_2})^{\frac{1}{a_1-a_2}}-g_3(t)}, \label{lvcase2}
\end{eqnarray}
where $g_3(t)=(a_1b_{22} e^{a_2t}+a_2b_{11}I_1 e^{a_1t})$. Depending on the signs and magnitudes of the system parameters $a_1,a_2,b_{11},b_{12}$ and $b_{22}$, the above solutions describe normalized interacting populations which asymptotically decay or grow or saturate.

Similarly for all the other integrable cases identified in this section, one
can derive the general solutions which are physically and mathematically relevant, often after some manipulations.
The details will be presented elsewhere.

\subsection{Application to non-autonomous system of first order ODE}
\label{sec4:4:2} As we mentioned in the introduction, one of our
motivations is to show that the procedure developed in \S3 is also
applicable to non-autonomous systems as well.

\subsubsection{Example 1: Complex Riccati equation}
To demonstrate this
point in brief let us consider the following first order
non-autonomous equations
\begin{eqnarray}
\dot{x}=\alpha_1(t)x-\alpha_2(t)y+\beta_1(t)(x^2-y^2)-2\beta_2(t)xy=\phi_1(t,x,y),\nonumber\\
\dot{y}=\alpha_1(t)y+\alpha_2(t)x+\beta_2(t)(x^2-y^2)+2\beta_1(t)xy=\phi_2(t,x,y),
\label{cfn02}
\end{eqnarray}
where $\alpha_i(t)$ and $\beta_i(t)$, $i=1,2$, are arbitrary functions of $t$. Equation (\ref{cfn02}) describes the dynamics of two interacting species with time modulated parameters. This coupled equation is essentially the real form of the complex Riccati/Bernoulli
equation $\dot{z}=\alpha(t)z+\beta(t)z^2$, where $z=x+iy$, $\alpha(t)=\alpha_1(t)+i\alpha_2(t)$ and $\beta(t)=\beta_1(t)+i\beta_2(t)$.
Substituting the form of $\phi_i$'s, $i=1,2,$ into the
determining equation (\ref{eq10}) and solving the latter  we obtain
the following forms for $S$ (generalizing the ansatz (\ref{eq18})), namely
\begin{eqnarray}
S_1=\frac{(y^2-x^2)\cos(\omega(t))-2xy\sin(\omega(t))}
{(x^2-y^2)\sin(\omega(t))-2xy\cos(\omega(t))},\nonumber\\
S_2=\frac{(y^2-x^2)\sin(\omega(t))+2xy\cos(\omega(t))}
{(y^2-x^2)\cos(\omega(t))-2xy\sin(\omega(t))}, \label{cfn03}
\end{eqnarray}
where $\omega(t)=\int\alpha_2(t) dt$.
Now inserting the above forms of $S_1$ and $S_2$ into equation~(\ref{eq11}) and
solving the resultant equations we obtain the following integrating factors, that is,
\begin{eqnarray}
K_1=\frac{e^{\int\alpha_1(t) dt}}{(x^2+y^2)^2}\bigg((y^2-x^2)\sin(\omega(t))+2xy\cos(\omega(t))\bigg),\nonumber\\
K_2=\frac{e^{\int\alpha_1(t) dt}}{(x^2+y^2)^2}\bigg((x^2-y^2)\sin(\omega(t))+2xy\cos(\omega(t))\bigg).\label{cfn04}
\end{eqnarray}
Substituting the complete sets $(R_i(=S_iK_i),K_i)$, $i=1,2$, into
equation~(\ref{eq09}) and evaluating the integrals we obtain
\begin{eqnarray}
I_1=\frac{e^{\int\alpha_1(t) dt}}{(x^2+y^2)}\bigg(x\cos(\omega(t))+y\sin(\omega(t)\bigg)
+\gamma_1(t), \nonumber\\
I_2=\frac{e^{\int\alpha_1(t) dt}}{(x^2+y^2)}\bigg(x\sin(\omega(t))-y\cos(\omega(t))\bigg)
+\gamma_2(t),\label{cfn05}
\end{eqnarray}
where
\begin{eqnarray}
\gamma_1(t)=\int
\bigg(\beta_1(t)\cos(\omega(t))-
\beta_2(t)\sin(\omega(t))\bigg)e^{\int\alpha_1(t)dt}dt, \nonumber\\
\gamma_2(t)=\int
\bigg(\beta_2(t)\cos(\omega(t))+
\beta_1(t)\sin(\omega(t))\bigg)e^{\int\alpha_1(t)dt}dt. \label{cfn06}
\end{eqnarray}
From the integrals $I_1$ and $I_2$, the general solution for the
equation (\ref{cfn02}) can be fixed easily in the form
\begin{eqnarray}
x(t)=\frac{e^{\int\alpha_1(t) dt}}{\gamma_3(t)}
\bigg((I_1-\gamma_1(t))\cos(\omega(t))+(I_2-\gamma_2(t))\sin(\omega(t))\bigg),
\nonumber\\
\quad y(t)=\frac{e^{\int\alpha_1(t) dt}}{\gamma_3(t)}
\bigg((I_1-\gamma_1(t))\sin(\omega(t))-(I_2-\gamma_2(t))\cos(\omega(t))\bigg),
\label{cfn07}
\end{eqnarray}
where $\gamma_3(t)=(I_1-\gamma_1(t))^2+(I_2-\gamma_2(t))^2$. Again depending on the nature of the system parameters, the above solutions represent oscillatory or decaying or growing populations.

\subsubsection{Example 2: A predator-prey equation}
To demonstrate the theory for non-autonomous systems further, we consider another example which is a non-autonomous predator-prey equation,
\begin{eqnarray}
\dot{x}=\alpha_1 x+\gamma_1 e^{-\beta_1 t}xy=\phi_1,\quad
\dot{y}=\alpha_2 y+\gamma_2 e^{-\beta_2 t}xy=\phi_2, \label{2ex201}
\end{eqnarray}
where $\alpha_i,\;\beta_i$ and $\gamma_i$, $i=1,2,$ are arbitrary
parameters. Substituting the form of $\phi_i$'s, $i=1,2,$ into the
determining equation (\ref{eq10}) and solving the latter  we obtain
the following form for $S$ for the parametric choice
$\beta_1=\beta_2=\beta$, namely
\begin{eqnarray}
S_1=-\frac{(\alpha_2-\beta)y+\gamma_2 e^{-\beta
t}xy}{(\alpha_1-\beta) x+\gamma_1 e^{-\beta t}xy}. \label{2ex202a}
\end{eqnarray}
Now inserting the above form of $S_1$ into equation~(\ref{eq11}) and
solving the resultant equation we obtain
\begin{eqnarray}
K_1=\frac{(\alpha_1-\beta)+\gamma_1 e^{-\beta t}y}{y}.
\label{2ex202b}
\end{eqnarray}
Substituting the forms $R_1(=S_1K_1)$ and $K_1$ into
equation~(\ref{eq09}) and evaluating the integrals we obtain
\begin{eqnarray}
I_1=(\gamma_2x-\gamma_1y)e^{-\beta
t}+(\alpha_2-\beta)\log{x}-(\alpha_1-\beta)\log{y}-(\alpha_2-\alpha_1)\beta
t. \label{2ex202c}
\end{eqnarray}
Unfortunately we could not find a second integral in this case. However, for the further parametric restriction, namely $\alpha_1=\alpha_2=\alpha$
 we obtain $S_2$ of the form
\begin{eqnarray}
S_2=-\frac{\gamma_2}{\gamma_1}. \label{2ex202}
\end{eqnarray}
Substituting $S_2$ into equation~(\ref{eq11}) and solving
the resultant equation we obtain
\begin{eqnarray}
K_2=\gamma_1e^{-\alpha t}. \label{2ex203}
\end{eqnarray}
Plugging the forms $R_2(=SK)$ and $K_2$ into
equation~(\ref{eq09}) we obtain $I_2$ as
\begin{eqnarray}
I_2=(\gamma_2x-\gamma_1y)e^{-\alpha t}. \label{2ex204}
\end{eqnarray}
From the integrals $I_1$ and $I_2$, the general solution for the
equation (\ref{2ex201}) with $\alpha_1=\alpha_2=\alpha$ and $\beta_1=\beta_2=\beta$
can be fixed easily in the form
\begin{eqnarray}
x(t)=\frac{I_2e^{\alpha t}}
{(\gamma_2-\gamma_1e^{\frac{(I_1-I_2e^{(\alpha-\beta)t})}{\beta-\alpha}})},
\quad y(t)=\frac{I_2e^{\alpha
t}}{(\gamma_2e^{\frac{(I_1-I_2e^{(\alpha-\beta)t})}{\alpha-\beta}}-\gamma_1)}.
\label{2ex205}
\end{eqnarray}
It is obvious that depending upon the signs and magnitudes of the parameters $\alpha$ and $\beta$ the general solution either decays or grows or saturates in the asymptotic limit.

\section{Prelle-Singer procedure for three coupled first order ODEs}
\label{sec4:5}
Next, we focus our attention on a system of three coupled first order
ODEs of the form
\begin{eqnarray}
&&\dot{x}=\frac{P_1(t,x,y,z)}{Q_1(t,x,y,z)}, \quad
\dot{y}=\frac{P_2(t,x,y,z)}{Q_2(t,x,y,z)}, \quad
\dot{z}=\frac{P_3(t,x,y,z)}{Q_3(t,x,y,z)},\label {3eq01}
\end{eqnarray}
where $P_i$'s and $Q_i$'s, $i=1,2,3,$ are analytic functions in
$x,\;y$ and $z$ with coefficients in the field of complex numbers.
Further, we assume that the ODE (\ref{3eq01}) admits a first
integral $I(t,x,y,z)=C,$ with $C$ constant on the solutions so that
the total differential becomes
\begin{eqnarray}
dI=I_tdt+I_{x}dx+I_{y}dy+I_{z}dz=0. \label {3eq02}
\end{eqnarray}
Now let us rewrite the equation (\ref{3eq01}) in the form
\begin{eqnarray}
\frac{P_1}{Q_1}dt-dx=0,\quad
\frac{P_2}{Q_2}dt-dy=0,\quad
\frac{P_3}{Q_3}dt-dz=0. \label {3eq03}
\end{eqnarray}
Hence, on the solutions, the 1-forms (\ref{3eq02}) and (\ref{3eq03})
must be proportional. Multiplying the first, second and third
equations in (\ref{3eq03}) by the functions $R(t,x,y,z)$, $K(t,x,y,z)$
and $M(t,x,y,z)$, respectively,  which act as the integrating
factors of the corresponding equations, we have on the solutions that
\begin{eqnarray}
&&dI=(R\phi_1+K\phi_2+M\phi_3)dt-Rdx-Kdy-Mdz=0,\;\;\label {3eq04}
\end{eqnarray}
where $ \phi_i\equiv {P_i}/{Q_i},\; i=1,2,3$. Comparing equations (\ref{3eq04})
and (\ref{3eq02}) we have, on the solutions, the relations
\begin{eqnarray}
 I_t  =(R\phi_1+K\phi_2+M\phi_3),
 \quad  I_{x}  = -R, \quad I_{y} =-K, \quad I_{z} =-M.
  \label {3eq05}
\end{eqnarray}

The compatibility conditions between the equations
(\ref{3eq05}) provide us the following determining equations for the
integrating factors  $R,K$ and $M$:
\begin{eqnarray}
&&R_t+\phi_1R_x+\phi_2R_y+\phi_3R_z  =-(R\phi_{1x}+K\phi_{2x}+M\phi_{3x}),
\label {3eq06}\\
&&K_t+\phi_1K_x+\phi_2K_y+\phi_3K_z  =-(R\phi_{1y}+K\phi_{2y}+M\phi_{3y}),
\label {3eq07}\\
&&M_t+\phi_1M_x+\phi_2M_y+\phi_3M_z  =-(R\phi_{1z}+K\phi_{2z}+M\phi_{3z}),
\label {3eq08}\\
&&R_{y} =K_{x},
\quad R_{z} =M_{x},
\quad K_{z} =M_{y}. \label {3eq11}
\end{eqnarray}

On the other hand integrating equations~(\ref{3eq05}), we obtain the integral
of motion,
\begin{eqnarray}
I=r_1+r_2+r_3-\int\bigg[M+\frac{d}{dz}\bigg(r_1+r_2+r_3\bigg)\bigg]dz,
\label {3eq12}
\end{eqnarray}
where
\begin{eqnarray}
r_1&=&\int\bigg(R\phi_1+K\phi_2+M\phi_3\bigg)dt,
\qquad
r_2=-\int\bigg(R+\frac{d}{dx}(r_1)\bigg)dx\nonumber\\
r_3&=&-\int\bigg(K+\frac{d}{dy}(r_1+r_2)\bigg)dy.
\nonumber
\end{eqnarray}
Naturally, for the complete integrability of Eq. (\ref{3eq01}) we require
three independent integrals and
so three independent sets of integrating factors $(R_i, K_i, M_i),\;i=1,2,3$.
\section{Method of Solving Determining Equations}
\label{sec4:6}
The determining equations (\ref{3eq06})-(\ref{3eq11}) are more complicated than
the two dimensional case discussed in \S3 and so to simplify the determining
equations we introduce
the transformations
\begin{eqnarray}
R=SM \quad \mbox{and} \quad K=UM,
\label {3eq12a}
\end{eqnarray}
where $S$ and $U$ are functions of $t,x,y$ and $z$, so that the equations
(\ref{3eq06})-(\ref{3eq11}) become
\begin{eqnarray}
D[S]  &=&S(S\phi_{1z}+U\phi_{2z}+\phi_{3z})-(S\phi_{1x}+U\phi_{2x}+\phi_{3x}),
\label {3eq13}\\
D[U]  &=&U(S\phi_{1z}+U\phi_{2z}+\phi_{3z})-(S\phi_{1y}+U\phi_{2y}+\phi_{3y}),
\label {3eq14}\\
D[M]  &=&-M(S\phi_{1z}+U\phi_{2z}+\phi_{3z}), \label {3eq15}\\
M_{x} &=&SM_{z}+MS_{z},
\quad M_{y} =UM_{z}+MU_{z} , \label {3eq17}\\
U_{x}-S_{y} &=&SU_{z}-US_{z}, \label {3eq18}\;\;\;D =\frac{\partial}{\partial{t}}
+\phi_1\frac{\partial}{\partial{x}}
+\phi_2\frac{\partial}{\partial{y}}+\phi_3\frac{\partial}{\partial{z}}.
\end{eqnarray}
One may note that two of the determining equations are still in coupled
form and the transformations are natural extensions of the two
coupled case. Of course, one may also consider alternate
possibilities, that is, either $R=\hat{S}K$ and $M=\hat{U}K$ or
$K=\tilde{S}R$ and $M=\tilde{U}R$. However, such possibilities
again lead to same results.

In the two coupled case (\ref{eq01}), we divided our analysis into
two categories, vide \S3$\;a$ (time-independent integrals) and
\S3$\;b$ (time-dependent integrals). However, in the present case we
divide our analysis into five categories, that is, (i) $I_x=0$ and
$I_t,\;I_y,\;I_z\neq0$, (ii) $I_y=0$ and $I_t,\;I_x,\;I_z\neq0$,
(iii) $I_z=0$ and $I_t,\;I_x,\;I_y\neq0$, (iv) $I_t=0$ and
$I_x,\;I_y,\;I_z\neq0$ and (v) $I_t,\;I_x,\;I_y,\;I_z\neq0$. We
intend to proceed in this way because we observed the absence
of a dynamical variable in some integrals in certain specific
dynamical systems of the type (\ref{3eq01}). We try to identify
these cases first. In fact, proceeding in this way we are able to
formulate a condition on the system variables and if the given
system satisfies this condition one can conclude that the given
dynamical system has the integral without that respective variable.
Since we have four variables, $t,x,y$ and $z$ we consider each one
of the cases separately and treat none of the variables being absent
as the fifth independent case.

Since we are dealing with a system of coupled three first order ODEs,
the complete
integrability is guaranteed by the presence of 2 time-independent
integrals (whereupon the system can be reduced to a single
quadrature) or 3 time dependent ones (in which case the
solution can be obtained in an algebraic way
(Bountis \textit{et al.} 1984)). In the following we will search for
such integrals.

\subsection{Caes 1: $I_x=0$ and $I_t,\;I_y,\;I_z\neq0$}
\label{sec4:6:1} In the case $I_x=0$, we have $R=0$ (vide equation
(\ref{3eq05})) which in turn implies that (i) either $S=0$ and
$M\neq0$ or (ii) $S\neq0$ and $M=0$ as can be seen from
(\ref{3eq12a}). In the former case, $S=0,\;M\neq0$, one can easily
fix the form of $U$, from equations (\ref{3eq13}) and
(\ref{3eq18}) as
\begin{eqnarray}
U = -\frac{\phi_{3x}}{\phi_{2x}}, \quad U_x=0.\label {3eq19a}
\end{eqnarray}
On the other hand the choice $M=0$ and $S\neq0$ when $R=0$ leads to
the case where one of the dynamical variables become uncoupled (see
equations (\ref{3eq06})-(\ref{3eq11})) and
so effectively a system of two coupled first order ODEs results in,
which we have already discussed. So this choice is not considered
further. Inserting the above form (\ref{3eq19a}) into (\ref{3eq14})
we arrive at the condition
\begin{eqnarray}
\phi_{3x}(\phi_{2xt}+\phi_{2}\phi_{2xy}+\phi_{3}\phi_{2xz}
-\phi_{2z}\phi_{3x}-\phi_{2x}\phi_{2y})-\phi_{2x}(\phi_{3xt}
\nonumber\\\qquad
+\phi_{2}\phi_{3xy}+\phi_{3}\phi_{3xz}
-\phi_{2x}\phi_{3y}-\phi_{3x}\phi_{3z})=0.\label {3eq19aa}
\end{eqnarray}
The condition (\ref{3eq19aa}) gives us the integrable cases for which the
system possesses the integrals of motion with $I_x=0$. Now
substituting (\ref{3eq19a}) into (\ref{3eq15})
we obtain the following determining equation for $M$, that is,
\begin{eqnarray}
D[M]  &=&M(\frac{\phi_{3x}}{\phi_{2x}}\phi_{2z}-\phi_{3z}).\label {3eq20a}
\end{eqnarray}
Again to solve equation (\ref{3eq20a}) one has to make a suitable ansatz for
$M$. Choosing appropriate ansatz for $M$ and solving the equation (\ref{3eq20a})
one can get explicit form for $M$. Once $M$ is known, the integrating
factors can be fixed from the relations
$K=UM$ and $R=SM=0$.
Now plugging the forms of $R,\;K$ and $M$ into Eq.~(\ref{3eq12}) and
evaluating the integrals one can construct the integrals of motion
for the given system.
\subsection{Case 2: $I_y=0$ and $I_t,\;I_x,\;I_z\neq0$}
\label{sec4:6:2} The determining equations and conditions can be
fixed in a similar manner for this case, $I_y=0$ and $I_t,\;I_x,\;I_z\neq0$,
with the replacement $(S, U, M, \phi_{1}, \phi_{2}, \phi_{3}, x, y, z)$
$\rightarrow$ $(U, S, M, \phi_{2}, $ $\phi_{1}, \phi_{3}, y, x, z)$ in the
above analysis.

\subsection{Case 3: $I_z=0$ and $I_t,\;I_x,\;I_y\neq0$}
\label{sec4:6:3}
In the present case with the form of the integrating factors $R=SM$ and $K=UM$,
$I_z=0$ implies $M=0$ and so $R=0$ and $K=0$ as well, leading to an integral of motion which turns out to be constant. Therefore in this case we consider the other possibility
$R=\hat{S}K$ and $M=\hat{U}K$ and proceed as above. The final results are
obtained with the replacement $(S,U,M,\phi_{1},\phi_{2},\phi_{3},x,y,z)$
$\rightarrow$ $(\hat{U},\hat{S},K,\phi_{3},\phi_{1},\phi_{2},z,x,y)$ in case 1.

\subsection{Case 4: $I_t=0$ and $I_x,\;I_y,\;I_z\neq0$}
\label{sec4:6:4} Next, in the time independent case $I_t=0$ the
first equation in (\ref{3eq05}) gives
\begin{equation}
S=\frac{R}{M}=-\frac{(\phi_{3}+\phi_{2}U)}{\phi_{1}}.
\label{3it001}
\end{equation}
Substituting this form of $S$ into (\ref{3eq14}) and (\ref{3eq15})
we get the following form of determining equations for $U$ and $M$,
\begin{eqnarray}
D[U]  &=&\frac{\phi_{3}+\phi_{2}U}{\phi_{1}}
(U\phi_{1y}-\phi_{1z})+U(U\phi_{2z}+\phi_{3z}-\phi_{2y})-\phi_{3y}, \label {3it002}\\
D[M]  &=&M(\frac{\phi_{3}+\phi_{2}U}{\phi_{1}}\phi_{1z}-\phi_{2z}U
-\phi_{3z}). \label {3it003}
\end{eqnarray}
To solve the equations (\ref{3it002}) and (\ref{3it003}) we adopt the following methodology.

To start with, in order to solve (\ref{3it002}), we consider $U$ in the form
\begin{eqnarray}
U = \frac{A_1(x,y)+B_1(x,y)z}{A_2(x,y)+B_2(x,y)z},
\label{3it004}
\end{eqnarray}
where $A_i$'s and $B_i$'s, $i=1,2,$ are arbitrary functions of $x$ and
$y$. Substituting (\ref{3it004}) into (\ref{3it002}) and equating the
coefficients of different powers of  $z$ to zero, we get a set of determining
equations for the functions  $A_i$'s and $B_i$'s, $i=1,2,$. Solving these
determining equations we obtain explicit expressions of the functions $A_i$'s
and $B_i$'s, $i=1,2,$
and consequently the associated  function $U$.

Now substituting the forms of $U$ into equation~(\ref{3it003}) and
solving the resultant equation we obtain the corresponding form of
$M$. To solve the determining equation for $M$ we again seek the
ansatz of the form $M = \frac{U_d}{(A(x,y)+B(x,y)z)^r},$  where
$U_d$ is the denominator of $U$. Once $U$ and $M$ are fixed
then one has to verify the compatibility of this set $(S,U,M)$  with
the constraint equations~(\ref{3eq17})-(\ref{3eq18}). Now
substituting $R(=SK),\;K(=UM)$ and $M$'s into equation~(\ref{3eq12})
one can construct the  associated integrals. Finally, one can
proceed with a more generalized ansatz than (\ref{3it004}) if need
arises.

\subsection{Case 5: $I_t,\;I_x,\;I_y,\;I_z\neq0$ case}
\label{sec4:6:5} Solving the determining equations
(\ref{3eq13})-(\ref{3eq15}) are naturally more tedious with none of the
variables $(t,x,y,z)$ absent in $I$, when compared to the earlier cases. To
start with one may use the following simple ansatz to solve the
determining equations (\ref{3eq06})-(\ref{3eq08}), that is,
\begin{eqnarray}
R&=& A_1(t,x,y)+B_1(t,x,y)z,\qquad
K= A_2(t,x,y)+B_2(t,x,y)z,\nonumber\\
M&=& A_3(t,x,y)+B_3(t,x,y)z,
\label{3eq24}
\end{eqnarray}
where $A_i$'s and $B_i$'s $i=1,2,3,$ are
arbitrary functions of $t,x$ and $y$.
Depending on the nature of the equation (\ref{3eq01}), one may work
with more general forms like a rational one.
\section{Three coupled ODEs - Applications}
\label{sec4:7}
\subsection{Example: R$\ddot{o}$ssler system}
\label{sec4:7:1}
Let us consider the R$\ddot{o}$ssler system (Rossler 1976)
\begin{equation}
\frac{dx}{dt}=-y-z=\phi_{1},
\quad \frac{dy}{dt}=x+\alpha_1y=\phi_{2},
\quad \frac{dz}{dt}=\alpha_2+xz+\alpha_3z=\phi_{3},
\label{2tex01}
\end{equation}
where $\alpha_i$'s, $i=1,2,3,$ are arbitrary parameters. Several
works have been devoted to study the dynamics of this equation. Very
recently, Llibre \& Zhang (2002) and Zhang (2004) have studied
equation (\ref{2tex01}) using the so called Darboux method and
obtained conditions for integrability. We now apply our above method
to system (\ref{2tex01}) and explore new integrals, if they exist.

\subsubsection{$I_x=0$ and $I_t,\;I_y,\;I_z\neq0$}
\label{sec4:7:1:1}
Substituting (\ref{2tex01}) into (\ref{3eq19aa}) we get
\begin{equation}
z\alpha_1+\alpha_2=0
\label{2tex02}
\end{equation}
From equation (\ref{2tex02}) we conclude that $\alpha_1=\alpha_2=0$ so that
from (\ref{3eq19a}) we get $U = -z$.
The determining equation for $M$ turns out to be
\begin{eqnarray}
M_t+xM_y+z(x+\alpha_3)M_z&=&-M(x+\alpha_3),\label {2tex04}
\end{eqnarray}
in which we have taken $M_x=0$ (since $I_x=0$).
A simple solution for (\ref{2tex04}) is $M=1/z$.
Making use of the explicit forms of $U$ and $M$ and with the
parametric restriction $\alpha_1=\alpha_2=0$ we conclude that
$R=0,\; K =-1,\; M = \frac{1}{z}$.
Now substituting the functions $R,\;K$ and $M$ into equation~(\ref{3eq12}) and
evaluating the integrals we
obtain the following integral of motion
\begin{eqnarray}
I=y+\alpha_3t-\log(z).\label{2tex08}
\end{eqnarray}
The integral (\ref{2tex08}) with $\alpha_3=0$ has already been given by
Llibre \& Zhang (2002) and Zhang (2004).  The
integral (\ref{2tex08}) with $\alpha_3\neq0$ is {\it new to the literature} at least
to our knowledge.
\subsubsection{$I_y=0,\;I_t,\;I_x,\;I_z\neq0$ and
$I_z=0,\;I_t,\;I_x,\;I_y\neq0$}
\label{sec4:7:1:2}
Proceeding appropriately we could not find any integrable case in the
R$\ddot{o}$ssler system belonging to these categories.

\subsubsection{$I_t=0$ and $I_x,\;I_y,\;I_z\neq0$}
\label{sec4:7:1:4}
In this case the function $S$ can be fixed in the form (vide equation
(\ref{3it001}))
\begin{equation}
S=\frac{\alpha_2+(x+\alpha_3)z+(x+\alpha_1y)U}{y+z}.
\label{2tex11}
\end{equation}
Substituting (\ref{2tex01}) into (\ref{3it002}) we get
\begin{eqnarray}
D[U]  &=&\frac{\alpha_2+(x+\alpha_3)z+(x+\alpha_1y)U}{y+z}(U-1)
+(x+\alpha_3-\alpha_1)U.
\label {2tex12}
\end{eqnarray}
Substituting (\ref{3it004}) into (\ref{2tex12}) and solving the
resultant equations we obtain non trivial forms of $U$ for the
specific  parametric restriction $\alpha_1=\alpha_2=\alpha_3=0$ and
then making use of the $U$ forms into (\ref{2tex11}) we obtain
$(S_1,U_1)=(x,y),\;(S_2,U_2)=(0,-z)$. Now substituting the forms of
$U$ into  equation~(\ref{3it003}) and solving the resultant equation
we obtain $M_1=-1$ and
$M_2=-e^{-y}$. Now inserting the integrating factors
$R_i$'s$(=S_iM_i),\;K_i$'s$(=U_iM_i)$ and $M_i$'s$,\;i=1,2,$ into
(\ref{3eq12}) and evaluating the integrals we arrive at the expressions
\begin{eqnarray}
I_1=x^2+y^2+2z, \quad
I_2=ze^{-y}.
\label {2tex15}
\end{eqnarray}
These two integrals are already known (Llibre \& Zhang
2002; Zhang 2004).
\subsubsection{$I_t,\;I_x,\;I_y,\;I_z\neq0$}
\label{sec4:7:1:5}
In this case, substituting $\phi_i$'s, $i=1,2,3,$ into
(\ref{3eq06})-(\ref{3eq08}) and solving the resultant system of equations
with the ansatz (\ref{3eq24}) we
obtain  the integrating factors for the parametric choice
$\alpha_1=\alpha_3=0$, that is, $R = -x,\quad K = -y,\quad M = -1$
and the corresponding integral of motion takes the form
\begin{eqnarray}
I=(x^2+y^2+2z-2\alpha_2t).
\label{2tex31}
\end{eqnarray}
The integral (\ref{2tex31}) has also been reported by Zhang (2004).

We conclude this section by mentioning that our studies reveal that
the system (\ref{2tex01}) possesses a time dependent integral for
the parametric choice $\alpha_3\neq0$.

\subsection{Example:2 3D-LV system}
\label{sec4:7:2}
Let us consider a 3D LV model for competition
between three populations whose dynamical evolution is determined by
the following equations (Cairo 2000)
\begin{eqnarray}
\dot{x}&=&x(\alpha_{1}+a_1 x+b_1 y+c_1 z),\qquad
\dot{y}=y(\alpha_{2}+a_2 x+b_2 y+c_2 z),\nonumber\\
\dot{z}&=&z(\alpha_{3}+a_3 x+b_3 y+c_3 z) \label{ml:01}
\end{eqnarray}
where $\alpha_i,a_i,b_i$ and $c_i$, $i=1,2,3,$ are arbitrary
parameters. Needless to say 3D-LV system is one of the challenging
problems and a testing ground for several analytical methods.
In the following we identify the integrals of motion for
certain specific parametric choices in (\ref{ml:01}) using the
procedure given above. For the purpose of
demonstration, in the following we present our results only for a couple of cases
and a detailed analysis will be presented separately.

\subsubsection{case 1:}
\label{sec4:7:2:1}
For the parametric choice $a_i=a$, $b_i=b$ and $c_i=c$, $\;i=1,2,3$,
we find the following three complete sets of integrating factors
$(R_i,K_i,M_i)$, $i=1,2,3,$
\begin{eqnarray}
&&R_1=0,\qquad K_1 = \frac{e^{(\alpha_3-\alpha_2)t}z}{y^2},\qquad
M_1 = -\frac{e^{(\alpha_3-\alpha_2)t}}{y},\nonumber\\
 &&R_2 = \frac{e^{(\alpha_3-\alpha_1)t}z}{x^2},\qquad K_2 = 0,\qquad
M_2 = -\frac{e^{(\alpha_3-\alpha_1)t}}{x},\nonumber\\
&&R_3=-\frac{e^{\alpha_1t}h(y,z)}{x^2},\; K_3 = \alpha_3
\frac{e^{\alpha_1t}}{ax},\; M_3 = \alpha_2 \frac{e^{\alpha_1t}}{ax},
\label{ml:02}
\end{eqnarray}
where $h(y,z)=(\alpha_2\alpha_3+ b\alpha_3y+ c\alpha_2 z)$. Now
substituting the functions $R,\;K$ and $M$ into
equation~(\ref{3eq12}) one can obtain the following integrals of
motion
\begin{eqnarray}
I_1&=&\frac{e^{(\alpha_2-\alpha_3)t}z}{y},\quad
I_2=\frac{e^{(\alpha_1-\alpha_3)t}z}{x},\nonumber\\
I_3&=&\frac{e^{\alpha_1t}(\alpha_1\alpha_2\alpha_3+\alpha_2\alpha_3ax
+\alpha_1\alpha_3 b y+\alpha_1\alpha_2 cz)}{x}. \label{ml:03}
\end{eqnarray}
From the integrals $I_1,I_2$ and $I_3$, we can deduce the general solution
for the equation (\ref{ml:01}) for the parametric choice $a_i=a$, $b_i=b$ and
$c_i=c$, $\;i=1,2,3$, of the form
\begin{eqnarray}
x(t)&=&\frac{\alpha_1\alpha_2\alpha_3 I_1 e^{\alpha_1t}}
{I_3I_1-(a\alpha_2\alpha_3 I_1 e^{\alpha_1t}
+b\alpha_1\alpha_3I_2 e^{\alpha_2t}+c\alpha_1\alpha_2I_1I_2e^{\alpha_3t})},\nonumber\\
y(t)&=&\frac{\alpha_1\alpha_2\alpha_3 I_2 e^{\alpha_2t}}{I_3I_1-(a\alpha_2\alpha_3 I_1 e^{\alpha_1t}
+b\alpha_1\alpha_3I_2 e^{\alpha_2t}+c\alpha_1\alpha_2I_1I_2e^{\alpha_3t})},\nonumber\\
z(t)&=&\frac{\alpha_1\alpha_2\alpha_3I_1I_2e^{\alpha_3t}} {I_3I_1-(a\alpha_2\alpha_3 I_1 e^{\alpha_1t}
+b\alpha_1\alpha_3I_2 e^{\alpha_2t}+c\alpha_1\alpha_2I_1I_2e^{\alpha_3t})}. \label{ml:04}
\end{eqnarray}
For the above choice of parameter the system (\ref{ml:01}) is a
completely integrable one.

\subsubsection{case 2:}
\label{sec4:7:2:3}
For the parametric
choice $\alpha_i=\alpha,\;i=1,2,3$, $a_3/3=-a_2/2=a_1$,
$c_1=-c_3,\;b_1=-b_3$ and $c_2=b_2=0$ we obtain two complete sets
of integrating factors of the form
\begin{eqnarray}
&&R_1=e^{-4\alpha t}y^2z,\;\;\; K_1 = 2e^{-4\alpha
t}xzy,\;\;\;
M_1 =e^{-4\alpha t}xy^2,\nonumber\\
 &&R_2 =a_1\alpha y(6a_1x-3c_1z+2b_1(y-\frac{c_1}{\alpha}z))e^{-3\alpha
t},
 \nonumber\\
 && K_2 =\alpha(3a_1^2x^2+b_1^2y^2+a_1x(4b_1y-c_1z(3+\frac{4}{\alpha}b_1y))
e^{-3\alpha t},\nonumber\\
 && M_2 =-a_1c_1xy(3\alpha+2b_1y)e^{-3\alpha t}. \label{ml:07}
\end{eqnarray}
Now substituting the integrating factors $R,\;K$ and $M$ into
equation~(\ref{3eq12}) one can obtain the corresponding integrals of
motions, namely
\begin{eqnarray}
&&I_1=e^{-4\alpha t}xy^2z,\nonumber\\
 &&I_2=e^{-3\alpha
t}\bigg((\alpha a_1xy(3c_1z-3a_1x-2b_1y+\frac{2}{\alpha}b_1c_1yz)-\frac{1}{3}b_1^2y^3\bigg).
\label{ml:08}
\end{eqnarray}
We could not find the third integrating factor $(R_3,K_3,M_3)$ within our
ansatz, (\ref{3eq24}), and further detailed exploration is needed to
conclude whether it exists or not to find the third integral if it exists.

\section{Prelle-Singer procedure for $n$ ($>$3) coupled first order ODEs}
\label{sec4:8}
The above procedure to find integrating factors and integrals can be extended in principle to a system of $n$ coupled first order ODEs $(n>3)$. In this cases, we get $n$ determining equations for the $n$ integrating factors along with $n(n-1)/2$ constraint equations, which one can solve algorithmically by following the above procedure. Further, the same procedure can also be applied in principle to any higher order as well as coupled higher order equations. This is because, any higher order equation can always be rewritten equivalently as a system of first order ODEs. For example, the second order equation $\ddot{x}=f(\dot{x},x,t)$ can be written in the first order form as $\dot{x}=y,\;\dot{y}=f(y,x,t)$. Then the determining equations for this first order form, namely equations (\ref{eq10})-(\ref{eq12}) can be related to the determining equations (2.6)-(2.8) given by us earlier
(Chandrasekar \textit{et al.} 2005)
for the second order form $\ddot{x}=f(\dot{x},x,t)$. However, one has to note that each one of the procedures has its own merits and demerits, for example in constructing nonstandard Hamiltonian structures. Similarly, a straightforward first order form for the third order equation $\dddot{x}=f(\ddot{x},\dot{x},x,t)$ is $\dot{x}=y,\;\dot{y}=z,\;\dot{z}=f(x,y,z,t)$. In this case also the determining equations for the first order form can be related to that of the third order ODEs as given by us earlier 
(Chandrasekar \textit{et al.} 2006). Again analyzing third order equations as such has its own practical advantages. This analogy can also be extended to coupled higher order equations as well in principle, though in the actual analysis one may have to make a judicious choice of which one of the methods is advantageous for investigating the integrability aspects.

\section{Linearization}
\label{sec4:9}
In this section, we describe a procedure to deduce the linearizing
transformations from the known integrals and illustrate the theory
with an example.

Let us assume that equation (\ref{eq01}) admits the following integral
\begin{eqnarray}
I=F(x,y,t).\label {modlin04}
\end{eqnarray}
Now  let us split the function $F_1$ in the form,
\begin{eqnarray}
I=F_1\left(\frac{1}{G_2(t,x,y)}\frac{d}{dt}G_1(t,x,y)\right).
\label{modlin06}
\end{eqnarray}
Now we identify the function $G_1$ as the new dependent variable
and the integral of $G_2$ over time as the new independent variable,
that is,
\begin{align}
w = G_1(t,x,y),\quad \tau = \int_o^t G_2(t',x,y) dt',
\label{modlin07}
\end{align}
We note here that the integration on the right hand side of
(\ref{modlin07}) leading to $\tau$ can be performed provided the function
$G_2$ is an exact derivative of $t$, that is,
$G_2=\frac{d}{dt}\tau(t,x,y)=\dot{x}\tau_x+\dot{y}\tau_y+\tau_t$.
In terms of the new variables,
equation (\ref{modlin06}) can be modified to the form
\begin{eqnarray}
I=F_1\left(\frac {dw}{d\tau}\right). \label{modlin08}
\end{eqnarray}
Inverting the relation (\ref{modlin08}) suitably, one can obtain a
linear equation,
\begin{eqnarray}
\frac {dw}{d\tau}=\hat{I}, \label{modlin09}
\end{eqnarray}
where $\hat{I}$ is a constant.  Or equivalently
\begin{eqnarray}
\frac {dw}{d\tau}=u=\hat{I},\quad \frac{du}{d\tau}=0.  \label{modlin11}
\end{eqnarray}
Equation (\ref{modlin11}) is the corresponding linear equation of (\ref{eq01}).
From equations (\ref{modlin07}) and (\ref{modlin11})
we have the following linearizing transformation for (\ref{eq01}),
namely
\begin{eqnarray}
w = \hat{G}_1(t,x,y),\quad u=F(x,y,t),\quad
\tau = \int_o^t \hat{G}_2(t',x,y) dt'.\label {modlin12}
\end{eqnarray}
In  this case the
new variables $w,u$ and $\tau$ helps us to transform the given system of
coupled first order nonlinear ODEs into a linear system of coupled first
order ODEs which in turn leads to the solution by trivial integration.
The above procedure can also be extended to more than two coupled first order ODEs straightforwardly and we do not present the details here.

\subsection{Example 1:}
To illustrate the underlying ideas let us consider the two
dimensional LV system (\ref{2ex01}) with the specific parametric
choice, $b_{11}=b_{21}$ and $b_{12}=b_{22}$,  namely
\begin{eqnarray}
\dot{x}=x(a_1+b_{11}x+b_{22}y),\quad
\dot{y}=y(a_2+b_{11}x+b_{22}y). \label{nmodlin01}
\end{eqnarray}
Let us consider the following first integral for equation
(\ref{nmodlin01}), namely
\begin{eqnarray}
I=\frac{y}{x}e^{(a_1-a_2)t}.\label {nmodlin04}
\end{eqnarray}
Now rewriting equation (\ref{nmodlin04}) using (\ref{nmodlin01}) in the form
(\ref{modlin06}) we get
\begin{eqnarray}
I=-\frac{1}{b_{22}}e^{-a_2t}\frac{d}{dt}\bigg[\bigg(\frac{1}{x}+\frac{b_{11}}{a_1}\bigg)e^{a_1t}\bigg].
\label{nmodlin06}
\end{eqnarray}
Then
\begin{align}
w = \bigg(\frac{1}{x}+\frac{b_{11}}{a_1}\bigg)e^{a_1t},\quad
\tau=-\frac{b_{22}}{a_2}e^{a_2t}. \label{nmodlin07}
\end{align}
From equations (\ref{nmodlin07}), (\ref{nmodlin04}) and
(\ref{modlin11}) we have the following linearizing transformation
for (\ref{nmodlin01}), namely
\begin{eqnarray}
w =\bigg(\frac{1}{x}+\frac{b_{11}}{a_1}\bigg)e^{a_1t},\quad
u=\frac{y}{x}e^{(a_1-a_2)t},\quad
\tau=-\frac{b_{22}}{a_2}e^{a_2t}.\label {nmodlin12}
\end{eqnarray}
In  this case the new variables $w,u$ and $\tau$ helps us to
transform the given system of coupled first order nonlinear ODE,
(\ref{nmodlin01}), into a linear system of coupled first order ODEs
of the form (\ref{modlin11}). The general solution can then be
deduced straightforwardly.

\subsection{Example 2:}
Similarly, for the specific parametric choice $a_1=a_2=0$,
$b_{12}=-b_{22}$ and $b_{21}=3b_{11}$, equation (\ref{2ex01}),
that is,
\begin{eqnarray}
\dot{x}=x(b_{11}x+b_{12}y),\quad \dot{y}=y(3b_{11}x-b_{12}y),
\label{emodlin01}
\end{eqnarray}
can be transformed to linear equation of the form (\ref{modlin11})
by the following linearizing transformations, namely (for the integral
see equation (\ref{3ex05b}))
\begin{eqnarray}
w
=\frac{1}{2}t^2-\frac{(b_{12}y+b_{11}x)}{2b_{11}x(b_{12}y-b_{11}x)^2},\quad
u=t+\frac{1}{b_{11}x-b_{12}y},\quad \tau =t.\label {emodlin12}
\end{eqnarray}

\section{Conclusion}
\label{sec4:10} In this paper, we have modified the Prelle-Singer
procedure such that it is applicable to both autonomous as well as
non-autonomous system of coupled first order ODEs. We have also
developed systematic procedures for finding both time independent
and time dependent integrals for them. From this analysis we have
answered the following open questions (i) How can the  PS method be
extended to non-autonomous system of coupled first order ODEs? (ii)
How to find the second or the time dependent integrals for the given
coupled first order ODEs and (iii) How can this procedure be
generalized to higher dimensions in order to find first integrals?
We have also shown that the determining equations for the time
independent integral in the two coupled equations of our method
coincides with the determining equations derived by Prelle and
Singer in their original paper. We have illustrated this procedure with
physically interesting examples, namely two dimensional
LV system, R$\ddot{o}$ssler system and 3D-LV system and
identified several integrable cases. Further, we have developed a
linearization procedure for coupled first order ODEs. Finally, we note that the procedures which we have developed in this paper, namely both the extended PS procedure and linearization can also be extended to any number of coupled first order ODEs.


The work of MS forms part of a research project sponsored by National Board for
Higher Mathematics, Government of India.  The work of ML forms part of a
Department of Science and Technology, Government of India sponsored research
project and is supported by a DST Ramanna Fellowship.

\end{document}